\documentclass{bioinfo}
\copyrightyear{2017} \pubyear{2017}
\usepackage{multirow}
\usepackage[table, x11names]{xcolor}
\usepackage{listings}
\usepackage{url}
\usepackage{color}
\access{Advance Access Publication Date: Day Month Year}
\appnotes{Manuscript Category}

\begin{document}
\firstpage{1}

\subtitle{Data and text mining}

\title[EBIC: an open source software for big and high-dimensional and big data biclustering analysis]{EBIC: an open source software for high-dimensional and big data biclustering analyses}

\author[Orzechowski \textit{et~al}.]
{Patryk Orzechowski\,$^{\text{\sfb 1,}\text{\sfb 2} *}$, Jason H. Moore\,$^{\text{\sfb 1,}*}$}
\address{
$^{\text{\sf 1}}$Institute for Biomedical Informatics, University of Pennsylvania, Philadelphia, PA 19104, USA, \\
$^{\text{\sf 2}}$Department of Automatics, AGH University of Science and Technology, al. Mickiewicza 30, 30-059 Krakow, Poland
}

\corresp{$^\ast$To whom correspondence should be addressed.}

\history{Received on XXXXX; revised on XXXXX; accepted on XXXXX}

\editor{Associate Editor: XXXXXXX}

\abstract{\textbf{Motivation:} In this paper we present an open source package with the latest release of EBIC, a next-generation biclustering algorithm for mining genetic data. The major contribution of this paper is adding full multi-GPU support, which makes it possible to run efficiently large genomic data mining analyses. Multiple enhancements to the first release of the algorithm include integration with R and Bioconductor, and an option to exclude missing values from analysis. \\
\textbf{Results:} EBIC was applied to datasets of different sizes, including a large DNA methylation dataset with 436,444 rows. For the largest dataset we observed over 6.6 fold speedup in computation time on a cluster of 8 GPUs compared to running the method on a single GPU. This proves high scalability of the method. \\
\textbf{Availability:} The latest version of EBIC could be downloaded from \url{http://github.com/EpistasisLab/ebic}. Installation and usage instructions are also available online.
\\
\textbf{Contact:} \href{patryk.orzechowski@gmail.com}{patryk.orzechowski@gmail.com}, \href{jhmoore@upenn.edu}{jhmoore@upenn.edu} \\
\textbf{Supplementary information:} Supplementary informations are available online.}
\maketitle

\section{Introduction}
Biclustering is an unsupervised machine learning technique which attempts to detect meaningful data patterns that are distributed across different columns and rows of the input dataset. This allows biclustering to capture heterogeneous patterns that manifest only in subsets of genes and subsets of samples. Biclustering has been commonly applied to genomic datasets \cite[]{Padilha2017} and has proven to be successful in revealing potential diagnostic biomarkers \cite[]{Liu2018}, or tumor transcription profiles in breast cancer \cite[]{Singh2018}.

With exponentially increasing sizes of the input datasets, there is an emerging need for effective and efficient methods that would scale well with growing amounts of data. Although there was discussion on possibility of applying biclustering to larger datasets \cite[]{Kasim2016,Padilha2017}, hardly any biclustering study involved large genomic dataset. This motivated emergence of parallel biclustering method. Some of the most recent parallel methods use multiple threads - e.g. runibic \cite[]{Orzechowski2018runibic}, or Message Passing Interface (MPI) - e.g. ParBiBit \cite[]{Gonzalez2018}, or GPU -  e.g. CCS \cite[]{Bhattacharya2017,Orzechowski2018ebic}. 

One of the recent advancements in biclustering area was introduction of EBIC - a parallel biclustering method, which takes advantage of multiple evolutionary computation strategies \cite[]{Orzechowski2018ebic}. This representative of hybrid biclustering algorithms \cite[]{Orzechowski2016data,Orzechowski2016text,Orzechowski2016pbba} has been shown to outperform multiple state-of-the-art methods in terms of accuracy. Although the original concept of EBIC provided theoretical support for multiple GPUs, all the previous evaluations have been made using a single GPU. Thus, the rationale of involving multiple GPUs was not clear. Another constraint for EBIC was hardware limitation of the size of the dataset to 65,535 rows per GPU. This required large clusters of GPUs in order to run analyses and greatly restricted application of the method.
 
In this paper we introduce the open source package built on top of the upgraded version of the method. First and foremost, a full support for multi-GPUs is added, which allows to analyze datasets with almost unlimited numbers of rows (available memory is a constraint). Secondly, the method has been integrated with Bioconductor, which enables the user to run all the analysis from the R level. Thirdly, a different method for performing analysis was added, which depends on the presence or absence of missing values within the data. Last, but not least, some bugs have been fixed and optimizations were made for more efficient memory management. All above combined make an this open source software ready out-of-the-box for big data biclustering analysis. 

%https://www.crcpress.com/Applied-Biclustering-Methods-for-Big-and-High-Dimensional-Data-Using-R/Kasim-Shkedy-Kaiser-Hochreiter-Talloen/p/book/9781482208238
%http://www.diss.fu-berlin.de/diss/servlets/MCRFileNodeServlet/FUDISS_derivate_000000010545/thesis_noCV.pdf
%DEBI
%ParBibit

\begin{methods}
\section{Methods}
The new version of EBIC provides a comprehensive open source framework for performing biclustering analysis. The major improvements over the original release of the method include:
\begin{itemize}
\item \textbf{Support for Big Data}. In the previous version of EBIC only a very limited number of rows could be processed on a single GPU. Kernel grid constrained the maximal number of rows analyzable by a EBIC to 65,535 per GPU. Thus, at least 8 GPUs were needed to analyze large datasets, e.g. modern methylation datasets. Our new implementation overcomes this limitation, allowing to analyze up to $2^{31}-1$ rows per single GPU (devices with computing capabilities 3.0). This greatly enhances the flexibility and applicability of the method to almost any type of data. This comes at a cost of reducing the size of genetic algorithm population down to 65,535. This remains a large number, as for the majority of genomic datasets the algorithm converged using a population size of 1,600 or less individuals.
\item \textbf{Handling missing values}. We introduce a very important feature which allows to remove the impact of missing values on the results of the method. As EBIC search is driven by counting of rows, a greater or equal relation between the values in columns used to capture missing values, instead of the real trends in the data. This posed a drawback, especially for datasets with high percentage of missing values. Instead of finding useful patterns in the data, EBIC used to become more attracted in detecting the emptiness. In the current release missing values might be replaced with a predefined value (e.g. 0 or 999), which is no longer counted towards the score of the bicluster. Thus, the method is focused on the real trends in the data, instead of emptiness.

\item \textbf{Different input file formats support}. EBIC allows different delimiters in input file. The data values might be separated by either comma, tabulator or semicolon. An upper left header is not required, which simplifies porting files between EBIC and R.

\item \textbf{Compatibility with R and Bioconductor}. The results returned by EBIC could be easily saved into a format loadable by Bioconductor R package \emph{biclust} in order to perform biological validation. In Supplementary Material we provide detailed workflow presenting how to use EBIC, all within R environment.

\item \textbf{Workflow for analysis of methylation data}. EBIC was capable to capture bio-meaningful signals in methylation data. A tutorial is presented in a Supplementary Material. 
\end{itemize}

\end{methods}

\section{Results}
In order to assess running times of the algorithm we have performed tests on from 1 up to 8 GPUs on datasets with varying number of rows and columns. The GEO accession numbers of the datasets as well as run times of the algorithm are presented in Table 1.

\begin{table}[ht]
\caption{Datasets used in the experiment as well as an average running time (in minutes) using a cluster of 8 GeForce GTX 1080 Ti GPUs.}
\label{gds-datasets}
\begin{center}
\begin{tabular}{r c c l r}
%\hline
\hline
\textbf{Dataset} &\textbf{Genes} &\textbf{Samples} &\textbf{Description} & \textbf{Run time}\\
\hline
GDS1490 &12,483&150& neural tissue profiling & 7.1 mins\\
GSE65194 & 54,675 &178 & breast cancer & 18.3 mins\\
GSE84493 & 436,444 &310& prostate cancer methylation & 24.5 mins \\
\end{tabular}
\end{center}
\end{table}

\begin{figure}[!htb]
\centering
\includegraphics[scale=0.5]{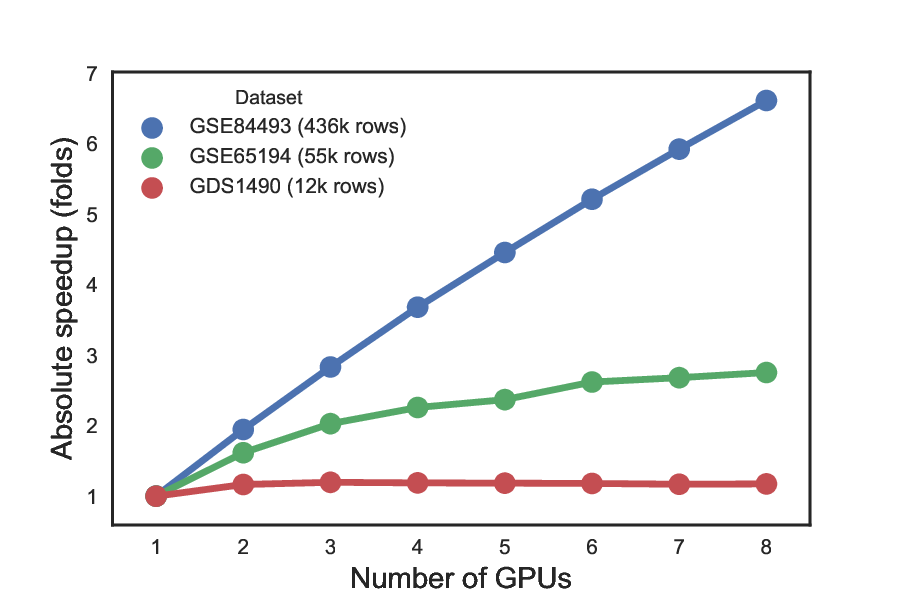}
\centering
\caption{Speedups obtained using multiple GPUs (GeForce GTX 1080 Ti) for the datasets from Table 1.}
\label{speedup}
\end{figure}

EBIC obtained up to 6.6x fold speedup using 8 GPUs on a dataset with over 436k rows. For the datasets with smaller number of rows the speedups were around 1.2x (12k rows) and 2.75x (55k rows). The relation between the different number of GPUs used and obtained speedup is presented in Fig. 1. 

\section{Conclusions}
In this paper we present the recent advancements in one of the leading biclustering methods. The algorithm was wrapped into a framework, which is conveniently integrated with R and allows multiple input file formats. In Supplementary Material we also demonstrate that even for such a large genomic dataset, the results provided by EBIC are bio-meaningful. We conclude that EBIC, released as open source package, is a very convenient tool for getting insight from large genomic datasets.

\section*{Funding}
This research was supported in part by PL-Grid Infrastructure and by grants LM012601, TR001263, ES013508 from the National Institutes of Health (USA).

\bibliographystyle{natbib}
\bibliography{main}

\end{document}